\documentstyle[epsf,osa,preprint]{revtex}
\tightenlines

\begin{document}

\preprint{\it Journal of the Optical Society
of America\/ \bf B 14\rm, 1273--1277  (1997)}

\title{Nonclassical interaction--free detection of
objects in a\\ monolithic total--internal--reflection resonator}

\author{Harry Paul$^{1}$ and Mladen 
Pavi\v ci\'c$^{1,2,\dag}$}

\address{$^1$Max--Planck--AG Nichtklassische Strahlung,
Humboldt University of Berlin,\\ D--12484 Berlin, Germany\\
$^2$Department of Mathematics, University of Zagreb, GF, 
Ka\v ci\'ceva 26, POB 217,\\ HR--10001 Zagreb, Croatia}

\maketitle
 
\begin{abstract} 
We show that with an efficiency exceeding 99\% one can use a
monolithic total--internal--reflection resonator in
order to ascertain the presence of an object without
transferring a quantum of energy to it. We also propose 
an experiment on the probabilistic meaning of the  electric 
field that contains only a very few photons. 
\end{abstract}

\bigskip 

\section{INTRODUCTION}
\label{sec:intro}Quantum optical measurements can exhibit an 
outstanding and completely nonclassical feature of detecting an 
object without transferring a single quantum of energy to it. For 
example, after the second beam splitter of a properly adjusted
Mach--Zehnder interferometer one can always put a detector in
such a position that it almost never detects a photon. If
it does, then we are certain that an object blocked one path
of the interferometer. A measurement which makes use of
this quantum mechanical feature was first designed by Elitzur
and Vaidman.\cite{bomb,vaid94} For dramatic effect
Elitzur and Vaidman assumed that the object is a bomb and
showed that it would explode in only 50\%\ of the tests
if asymmetrical beam splitter is employed. The tests, of
course, always have to be carried out with single photons.
Kwiat \it et al.\/\rm\cite{z95,zz95} proposed a set--up
which is based on ``weak repeated tests'' carried out by single
photons. They aim at reducing the probability of exploding the
bomb to as close to 0\%\ as possible and propose using two 
identical cavities weakly coupled by a highly reflective beam 
splitter. Due to the interference the probability for a photon
inserted into the first cavity to be located in it approaches 0, 
while the probability that it is found in the second cavity 
approaches 1, at a certain time $T_N$. However, if there
is an opaque object (a bomb) in the second cavity the 
probabilities reverse. So, if we insert a detector in the first 
cavity we almost never get a click if there is no absorber in 
the second cavity, and almost always if there is one. The 
probability of exploding the bomb when there is a detector in 
the first cavity approaches zero. The drawback of the proposal 
is that it is apparently very hard to carry it out. In 
particular, inserting a detector into a cavity at a given time 
and introducing a single photon into a cavity are, by no means, 
simple tasks.

In Sec.~\ref{sec:experiment} we propose a very simple,
very efficient, and easily feasible interaction--free 
experiment---with an arbitrary high probability of \it 
detecting the bomb without exploding it\/\rm---which is 
based on the resonance in a single monolithic
total--internal--reflection resonator which has recently
shown significant practical advantage and very high
efficiencies. The proposal assumes a pulse laser beam or a 
continuous wave (cw) laser beam and a properly cut isotropic 
crystal. The additional advantage of the present proposal 
over the proposals of Kwiat \it et al.\/\rm\cite{z95,zzz95} 
is that the latter ones infer information on
the presence or absence of the bomb in the system from
the absence of a detector click while we get the information 
from the firing of appropriate detectors.

Actually, we will perform a classical calculation. As is
well known, there is a formal correspondence between the
classical and the quantum description in the sense that
classical quantities, e.g., the amount of energy absorbed
by a bomb, are identical to quantum mechanical \it
ensemble averages. \rm When the absorbed energy is, on
average, only a small fraction of an energy quantum
$h\nu$, this means in reality that in most cases no
absorption takes place at all, and only in a few cases
one photon is actually absorbed. This is a consequence of
the corpuscular nature of light. In other words,
interpreting classical intensities as quantum mechanical
probabilities to find a photon, allows us to
``translate'' our classical results into quantum
mechanical predictions.

Our elaboration shows that, in accordance with quantum
theory, an overwhelming number of photons detected by a 
detector indicating a presence of a bomb in the system will 
not exchange any energy with the bomb. Only in rare cases a 
photon will transfer an energy quantum $h\nu$ to the bomb. 
Nevertheless, on average, there is no difference between the 
quantum and the classical picture. This means that many other 
parallels between the two pictures, such as wave amplitude 
calculations, coherence time and length, optical
resonator calculations, cavity photon decay time, etc.,
should be preserved. On the other hand, in one--photon quantum
interference the amplitudes of electric field have a 
probabilistic meaning only,  and one might wonder whether a 
classical reasoning would give a correct answer. For example, 
Weinfurter \it et al.\/\rm\cite{wzz95} and Fearn \it et
al.\/\rm\cite{frust} proposed an experiment which would decide 
whether sudden changing of boundary conditions of frustrated 
downconverted photons affects photons instantaneously or after 
a (classically untenable) delay which would allow all parts of 
the system (atoms emitting downconverted photons) to ``receive'' 
the information on the changed conditions.

In Sec.~\ref{sec:path} we propose an experiment
which would decide whether sudden changing of boundary
conditions imposed on photon paths (object--no object) redirect 
the photons (into $D_r$ instead into $D_t$ and \it vice 
versa\/\rm)  instantaneously (classically untenable) or after a 
delay which would allow for sufficiently many round trips to 
build up the interference.

\section{INTERACTION--FREE EXPERIMENT}
\label{sec:experiment}

The lay--out of the experiment is shown in Fig.~\ref{exp}.
The experiment uses an uncoated monolithic 
total--internal--reflection resonator (MOTIRR) coupled to 
two triangular prisms by the frustrated total internal 
reflection (FTIR). \cite{motirr1,motirr2}. A squared MOTIRR 
requires a relative refractive index with respect to the 
surrounding medium $n>1.41$ in order to confine a beam to 
the resonator (the angle of incidence being 45$^\circ$).
If, however, another medium (in our case the right triangular 
prism in Fig.~\ref{exp}) is brought within a distance of the
order of the wavelength, the total reflection within the 
resonator will be \it frustrated\/ \rm and a fraction of the 
beam will ``tunnel out'' from the resonator. Depending on the 
dimension of the gap and the polarization of the incidence 
beam one can well define reflectivity $R$ within the range 
from $10^{-5}$ to 0.99995. \cite{motirr2,ftir}
The main advantage of such a coupling---in comparison
with coated resonators---is that the losses are extremely 
small: down to 0.3\%. In the same way a beam can ``tunnel 
into'' the resonator through the left triangular prism in 
Fig.~\ref{exp}, provided the condition $n>1.41$ is fulfilled 
for the prism too.  The incident laser beam is chosen to be 
polarized perpendicularly to the incident plane so as to give 
a unique reflectivity for each photon. The faces of the 
resonator are polished spherically to give a large focusing 
factor. A round trip path for the beam is created in the
resonator as shown in Fig.~\ref{exp}. A cavity is cut in the 
resonator and filled with an index--matching fluid to reduce 
losses. (Before we carry out the measurement we have to wait 
until the fluid comes to a standstill to avoid a possible 
destabilization of the phase during round trips of the beam.) 
Now, if there is an object in the cavity in the round trip 
path of the beam in the resonator, the incident beam will be 
almost totally reflected (into $D_r$) and if there is no object, 
the beam will be almost totally transmitted (into $D_t$).

We start with a formal presentation of the experiment.
Our aim is first to determine the intensity of the beam reaching
detector $D_r$ when the crystal is at resonance, and then this
intensity when an opaque object (bomb) is in the round--trip path
in the crystal. A portion of a polarized incoming beam of
amplitude $A(\omega)$ is totally reflected from the inner surface 
of the left coupling prism (see Fig.~\ref{exp}) and suffers
a phase shift which ranges from $0^\circ$ (for the critical angle
$\theta_1=\arcsin n$) to $180^\circ$ (for $\theta_1=0^\circ$),
where $\theta_1$ is the incident angle at the prism inner
surface.\cite{wolf} When we couple the ring cavity (MOTIRR) to 
the prism, a part of the incoming beam will tunnel into MOTIRR 
and the reflected beam will suffer a new phase shift $\delta$ and 
the complex--field--reflection coefficient will read:
\begin{eqnarray}
re^{i\delta}=[1-{2\sin\delta_1\sin\delta_2\over \cosh 2bx
- \cos(\delta_1+\delta_2)}]\exp({\sin\delta_1\sinh
2bx\over\cos\delta_1
\cosh 2bx - \cos\delta_2})\,,
\label{eq:r1}
\end{eqnarray}
where $\delta_1$ and $\delta_2$ are the phase shifts of
the waves at
coupled surfaces of the prism and MOTIRR, respectively,
$x$ is the
gap between the prism and MOTIRR, and
$b={2\pi\over\lambda_0}\sqrt{n_1^2\sin^2\theta_1-n_2^2}$,
where
$n_1$ and $n_2$ are refraction indexes of the prism and
MOTIRR, respectively, and $\lambda_0$ is the vacuum wavelength. 
Thus detector $D_r$ rotated at the angle $\delta$ with respect 
to the incoming plane will receive the incoming power of the 
incident beam attenuated by $\vert r\vert ^2$.

At resonance the waves leaving the cavity after some
round trips in the cavity will add up
to a destructive interference so as to make the
reflected power vanish. Formally, it can be described as
follows. \cite{motirr2} The ratio of the reflected and
incoming powers is
\begin{eqnarray}
\eta=1-{c(x)\over 1+[{2{\cal F}\over\pi}\sin{\delta(x)+
\phi\over 2}]^2}\,,
\label{eq:power}
\end{eqnarray}
where $\phi$ is the total phase shift acquired by the wave during
one round trip, $\cal F$ is the finesse, and $c(x)$ is the 
coupling given by
\begin{eqnarray}
c(x)={(1-e^{-2\alpha})[1-\vert r(x)\vert ^2]\over[1-e^{-
\alpha}
\vert r(x)\vert]^2}\,,
\label{eq:c}
\end{eqnarray}
where $\alpha$ is a constant describing the round--trip losses.
At resonance the reflected power will vanish when MOTIRR is
\it impedance matched\/\rm, i.e., when the gap is adjusted so 
as to satisfy: $\vert r(x_m)\vert =e^{-\alpha}$. Then
$c(x_m)=1$ and $\phi=2N\pi - \delta$, where $N$ is an integer, 
so that $\eta=0$.

To understand this result we sum up the contributions
originating from round trips in the resonator, to the
reflected wave. The portion of the incoming beam of
amplitude $A(\omega)$ reflected into plane determined by 
$\delta$ is described by the amplitude 
$B_0(\omega)=-A(\omega)\sqrt{R}$, where $R=\vert r\vert ^2$ is
reflectivity. The transmitted part will travel around the 
resonator guided by one frustrated total internal reflection 
(at the face next to the right prism) and by two proper total 
internal reflections. After a full round trip the following 
portion of this beam joins the directly reflected portion of the 
beam by tunnelling into the left prism: 
$B_1(\omega)=A(\omega)\sqrt{1-R}\sqrt{R}\sqrt{1-R}\>e^{i\psi}$.
$B_2(\omega)$ contains three frustrated total internal
reflections and so on; each subsequent round trip contributes to 
a geometric progression which gives the reflected amplitude
\begin{eqnarray}
B_n(\omega)=A(\omega)\sqrt{R}\{-1+(1-
R)e^{i\psi}[1+Re^{i\psi}+
(Re^{i\psi})^2+\dots]\}=\sum^{n}_{i=0}B_i(\omega)\,,
\label{eq:total}
\end{eqnarray}
where $\psi=(\omega-\omega_{res})T$ is the phase added by each 
round trip. Here $\omega$ is the frequency of the incoming beam, 
$T$ is the round trip time, and $\omega_{res}$ is the selection 
frequency corresponding to a wavelength which satisfies
$\lambda=L/k$, where $L$ is the round trip length of the
cavity and $k$ is an integer. At summing up the round--trip
contributions we have taken into account that (because of the 
above condition imposed on the total phase shift $\phi$) all 
the contributions must lie in the reflected--wave plane and that 
their amplitudes must carry the opposite sign (to that of the 
reflected wave, $-A(\omega)\sqrt{R}$) so as to cancel out at 
resonance $\psi=0$.

To get a condensed insight into the physics of the experiment
let us first look at plane waves [$A(\omega)=A_0$]. The
limit of $B_n(\omega)$ yields the total amplitude of the 
reflected beam:
\begin{eqnarray}
B_r(\omega)=\lim_{n\rightarrow\infty}B_n(\omega)
=-A_0\sqrt{R}{1-e^{i\psi}\over1-R\,e^{i\psi}}
\,.
\label{eq:limit}
\end{eqnarray}
We see that for any $R<1$ and $\omega=\omega_{res}$, i.e., if
nothing obstructs the round trip of the beam, we get no 
reflection at all [i.e., no response from $D_r$ (see 
Fig.~\ref{exp})]. When a bomb blocks the round trip and $R$ is 
close to one, then we get almost a total reflection. In terms 
of single photons (which we can obtain by attenuating the 
intensity of a laser until the chance of having more
than one photon at a time becomes negligible) the probability 
of detector $D_r$ reacting when there is no bomb in the system 
is zero. A response from $D_r$ means an interaction--free 
detection of a bomb in the system. The probability of the 
response is $R$, the probability of making a bomb explode by 
our device is $R(1-R)$, and the probability of a photon 
exiting into detector $D_t$ is $(1-R)^2$.

A more realistic experimental approach we achieve by looking
at two possible sources of individual photons: a cw laser and a
pulse laser. For a pulse laser we make use of a Gaussian wave 
packet $A(\omega)=A\exp[-\tau^2(\omega-\omega_{res})^2/2]$,
where $\tau$ is the coherence time which obviously must be
significantly longer than the round trip time $T$. For a cw 
laser we use $A(\omega)=A\delta(\omega-\omega_{res})$, i.e.,
assume a well stabilized laser beam locked at $\omega_{res})$ 
with a negligibly small linewidth. Incident wave is described 
by:
\begin{eqnarray}
E^{(+)}_i(z,t)=\int^\infty_0A(\omega)e^{i(kz-\omega
t)}d\omega
\label{eq:e}
\end{eqnarray}
and the reflected wave by:
\begin{eqnarray}
E^{(+)}_r(z',t)=\int^\infty_0B(\omega)e^{i(kz'-\omega
t)}d\omega\,.
\label{eq:er}
\end{eqnarray}
Energy of the incoming beam is the energy flow integrated
over time:
\begin{eqnarray}
I_i=\int^\infty_{-\infty}E^{(+)}_i(z,t)E^{(-)}_i(z,t)dt=
\int^\infty_0 A(\omega)A^*(\omega)d\omega\,.
\label{eq:i}
\end{eqnarray}
Energy of the reflected beam is given analogously by
$I_{r;n}=\int^\infty_0 B_n(\omega)B_n^*(\omega)d\omega$.
Thus for the both types of lasers the ratio of energies $\eta$
as a function of the number of round trips $n$ is given by:
\begin{eqnarray}
\eta_n={I_{r;n}\over I_i}=R\{1-
{1-R\over1+R}[R^{2n}-1+2\sum^n_{j=1}(1+R^{2n-2j+1})R^{j-1}
\Phi(j)]\}\,,\label{eq:sum}
\end{eqnarray}
where $\Phi(j)=1$ for cw  lasers and
$\Phi(j)=\exp(-j^2a^{-2}4^{-1})$ for pulse lasers, where
$a\equiv \tau/T$. This expression is obtained by mathematical
induction from the geometric progression of the intensities of
the amplitudes given by Eq.~(\ref{eq:total}) and a subsequent
integration over wave--packets. It follows from 
Eq.~(\ref{eq:limit}) that the series with $\Phi(j)=1$ 
converges, so the series with
$\Phi(j)=\exp(-j^2a^{-2}4^{-1})$ converges as well.
For the latter $\Phi(j)$ a straight--forward calculation yields
\begin{eqnarray}
\lim_{n\rightarrow\infty}\eta_n={\int^\infty_0B_r(\omega)
B_r^*(\omega)d\omega\over
\int^\infty_0A(\omega)A^*(\omega)d\omega}=1-
(1-R)^2{\displaystyle\int_0^\infty{\displaystyle\exp[-
\tau^2(\omega-
\omega_{res})^2]d\omega\over\displaystyle1-
2R\cos[(\omega-
\omega_{res})\tau/a]+R^2}\over\displaystyle\int_0^\infty
\exp[-\tau^2(\omega-\omega_{res})^2]d\omega}
\,,\label{eq:eta}
\end{eqnarray}
where $B_r(\omega)$ is from Eq.~(\ref{eq:limit}). In 
Fig.~\ref{eff} the three upper curves represent three 
sums---obtained for three different values of $a$, 
respectively---that converge to values (shown as big dots) 
obtained from Eq.~(\ref{eq:eta}). The figure shows that $a$
and $n$ are closely related in the sense that the coherence 
length should always be long enough ($a>200$) to allow
sufficiently many round trips (at least 200).

A cw laser oscillating on a single transverse mode has the 
advantage of an excellent frequency stability (down to
10$\>$kHz, and with some effort even down to 1$\>$kHz, in the
visible range) and therefore a very long coherence length
(up to 300$\>$km). \cite{laser} This yields the zero intensity
at detector $D_r$ as with plane waves above. The only 
disadvantage of a cw laser is that we have to modify the setup 
by adding a gate which determines a time window
(1$\>$ms --- 1$\>\mu$s $<$ coherence time) within which the 
input beam arrives at the crystal and which allows the 
intensity in the cavity to build up. Then, the intensity of 
the beam should be lowered to make it probable for only one 
photon to appear within the time window. We start each testing 
by opening the gate, and when either $D_r$ or $D_t$ fires, or 
the bomb explodes, the testing is over. Of course, detectors 
might fail to react but this is not an essential problem 
because the single photon detector efficiency has already 
reached 85\%. For, this would result in a bigger time window,
but the chance of activating the bomb would remain very low.
In any case, the possibility of a 300$\>$km coherence length 
does not leave any doubt that a real experiment can be carried 
out successfully. It should be emphasized that we get
information on the presence or the absence of the bomb in
any case from a detector click, hence we need no
additional information that a photon has actually arrived
at the entrance surface. This is a great advantage over
the above--mentioned proposal by Kwiat et al. (1995) in
which the absence of the bomb is, in fact, inferred from
the absence of a detector click. When nothing happened
during the exposition time (due either to the absence of
a photon or detector inefficiency), the test has to be
repeated.

The main disadvantage of pulse lasers is that they have
mean frequency dependent on the working conditions of the 
laser, so each repetition of the experiment takes a 
considerable time to stabilize the frequency.
Their advantage is that they do not require any gates.

\section{VIRTUAL--OR--REAL--PATH EXPERIMENT}
\label{sec:path}
As we have seen in the previous section there is an
essential difference between our proposal and the one by 
Kwiat \it et al.\/\rm\cite{z95,zzz95} since their 
interaction--free experiment is based on \it repeated 
interrogations\/ \rm carried out by a ``real'' photon 
while ours is based on boundary conditions imposed on an 
``empty'' photon path which contains no photon.
Consequently, our approach does not give rise to the
quantum Zeno effect as opposed to theirs. \cite{zzz95}
Nevertheless, in our experiment we can formulate a question
whether the round trip path which a photon ``sees'' when
approaching the crystal is virtual or real. The question is 
similar to \it virtual--or--real--photon question\/ \rm 
posed by Weinfurter \it et al.\/\rm\cite{wzz95} and Fearn 
\it et al.\/\rm\cite{frust} Both questions assume answers
supported by classical formal reasoning and calculation.
In the latter experiment mirrors suppress propagation of
downconverted waves forming standing waves by reflecting
the waves back to the crystal. As soon as one removed the
mirrors and instantaneously replaced them by detectors the 
latter should fire triggered by the photons in the outgoing 
beams due to the changed boundary condition. In our 
experiment, changing the boundary conditions, i.e., 
switching on the round trip path, means allowing the round 
trip path to ``wind up'' (according to the calculation 
presented in Fig.~\ref{eff2}) even when there is no
single photon in the path: ``the paths are real.''

The experiment is presented in Fig.~\ref{exp2}. It is a
modification of the experiment shown in Fig.~\ref{exp}.
We tune in our FTIR--MOTIRR system so as to have as big
a gap between the coupling prisms and the crystal as 
possible (e.g., corresponding to $R=0.9999$). The Rochon 
prism \it p\/ \rm is rotated so as to fully match the 
phase shift as its O--wave. Therefore, when
the Pockels cell is \it off\/ \rm the round--trip path is
not influenced at all. When the Pockels cell is \it on\/ \rm
the path is redirected through Rochon prism $p$ (as E--wave)
into detector $D_p$. We switch on a cw laser and let it feed
the system. 

\parindent=0pt 
When the Pockels cell is \it on\/ \rm detector $D_r$ should 
fire with the probability approaching 1. When it is \it off\/ \rm  
detector should $D_t$ fire with the probability approaching 1. 

We carry out two kinds of measurement.
The first kind of measurement is switching the Pockels
cell \it on\/ \rm and monitoring $D_r$ immediately afterwards. 
We accommodate the intensity of the laser beam so as to have 
one photon in 0.1 ns in average. The fastest Pockels cells have 
reaction time down to 0.1 ns. The time an information traveling 
at the speed of light needs to spread from the Pockels cell to 
the incoming gap can be made as high as 4 ns by choosing the 
biggest available crystals. The fastest detectors have reaction 
time of under 1 ns. Before we switch on the Pockels cell almost 
only detector $D_t$ fires. After we switch on the Pockels cell 
we monitor detector $D_r$ and see whether it reacts 
instantaneously or after 4 ns.

The second kind of measurement is switching the Pockels
cell from \it on\/ \rm to \it off\/ \rm and monitoring detector
$D_r$ immediately afterwards. We lower down the intensity of 
the laser beam so as to have one photon in 10 ns in average.
We calculated that for $R=0.9999$ the resonance fully
establishes after 100 ns, i.e., after that time $D_r$ cannot 
fire (almost) at all. We monitor $D_r$ within this 100 ns and 
see whether detector $D_r$ stops firing immediately or only after
several firing within the first 100 ns. We have chosen 10 ns in
the incoming beam so as to make sure that after switching off
the Pockels cell only an ``empty'' wave is coming to the
system. A variety of the experiment would be to lower 
down the intensity of the laser beam further down to 
under one photon in 100 ns.

\section{CONCLUSION}
\label{sec:conclusion}

In summary, we have devised a feasible and, in principle,
rather simple experimental scheme for an
interaction--free detection of an absorbing object (bomb)
that rests on the classical behavior of a ring resonator
into which light is fed. In case of resonance and in the
absence of an obstacle within the cavity, a detector
$D_t$ in the exit channel will respond with very high
probability, whereas a detector $D_r$ placed in the light
beam reflected from the entrance mirror will almost never
detect a photon. When an object is introduced into the
cavity, the situation completely reverses. Hence the
absence or presence of the object will in any case be
indicated by a detector click. This is a great advantage
of our scheme, since there is no need to ensure that a
photon is actually impinging on the object. Moreover, we 
suggest a modification of the experimental scheme that would 
allow to measure time delays between blocking the resonator 
(or undoing a blocking) and the effect it has on the
detection probabilities. We apply the scheme on the 
Heisenberg microscope and the `Welcher Weg' experiment in 
Pavi\v ci\'c. \cite{pav} 

\bigskip\bigskip\bigskip
\bigskip

\acknowledgments
One of us (M.P.) gratefully acknowledges supports of the 
Alexander von Humboldt Foundation and of the Ministry of 
Science of Croatia. He would also like to thank 
David W.~Cohen, Department of Mathematics, Smith College, 
Northampton, Massachusetts for his comments on the 
manuscript. 

\bigskip\bigskip\bigskip
\bigskip\parindent=0pt

$^\dag$E-mail: mpavicic@faust.irb.hr; Web-page: http://m3k.grad.hr/pavicic.

\begin{figure}
\caption{Lay--out of the proposed experiment. For the shown free
round trips within the total--internal--reflection resonator the
incident laser beam tunnels in and out so as to give the zero 
intensity of the reflected beam, i.e., detector $D_r$ does not
react even when the incoming frustrated total reflectivity is
approaching one. However, when the bomb is immersed in the
(index--matching) liquid then practically the whole incoming beam 
reflects into $D_r$.}
\label{exp}
\end{figure}

\begin{figure}
\caption{Realistic values of $\eta$ [ratio of the incoming and
reflected powers, given by Eq.~(\protect\ref{eq:sum})] for 
$R=0.98$. For pulse lasers---3 upper curves represent sums from
Eq.~(\protect\ref{eq:sum}) [with $\Phi(j)=\exp(-j^2a^{-2}4^{-1})$]
as a function of $n$ for $a=100$, $a=200$, and $a=400$, where
$a\equiv \tau/T$ is a ratio of the coherence time $\tau$ and the
round--trip time $T$; dots represent the corresponding values of 
$\eta$ obtained from Eq.~(\protect\ref{eq:eta}). For cw 
lasers---the lowest curve represents the sum given by  
Eq.~(\protect\ref{eq:sum}) [with $\Phi(j)=1$] as a function of 
the number of round trips $n$.} \label{eff}
\end{figure}

\begin{figure}
\caption{Realistic values of $\eta$ [ratio of the incoming and
reflected powers, Eq.~(\protect\ref{eq:sum})] for $R=0.98$
(the lowest curve), 0.99, 0.995, 0.997, and 0.998 as functions 
of the number of round trips $n$.}
\label{eff2}
\end{figure}

\begin{figure}
\caption{Lay--out of the proposed virtual--or--real--path
experiment. When the Pockels cell $c$ is \it on\/ \rm it redirects
the round--trip path through Rochon prism $p$ into detector
$D_p$ and therefore almost only detector $D_r$ fires. When the 
Pockels cell $c$ is \it off\/ \rm there is no influence on the 
round--trip path and almost only detector $D_t$ fires.}
\label{exp2}
\end{figure}

\vfill\eject 

\epsffile{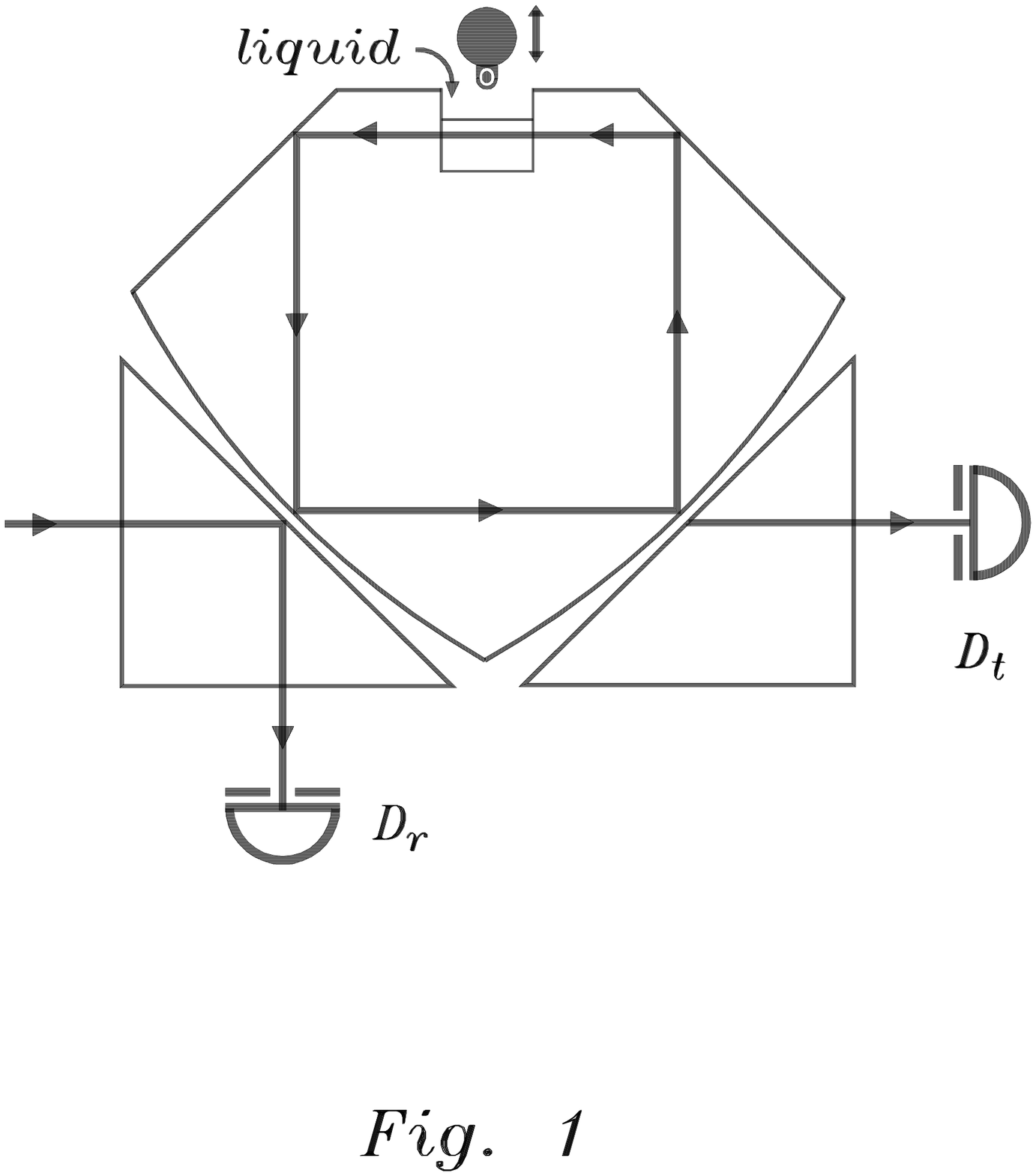}

\vfill\eject 

\epsffile{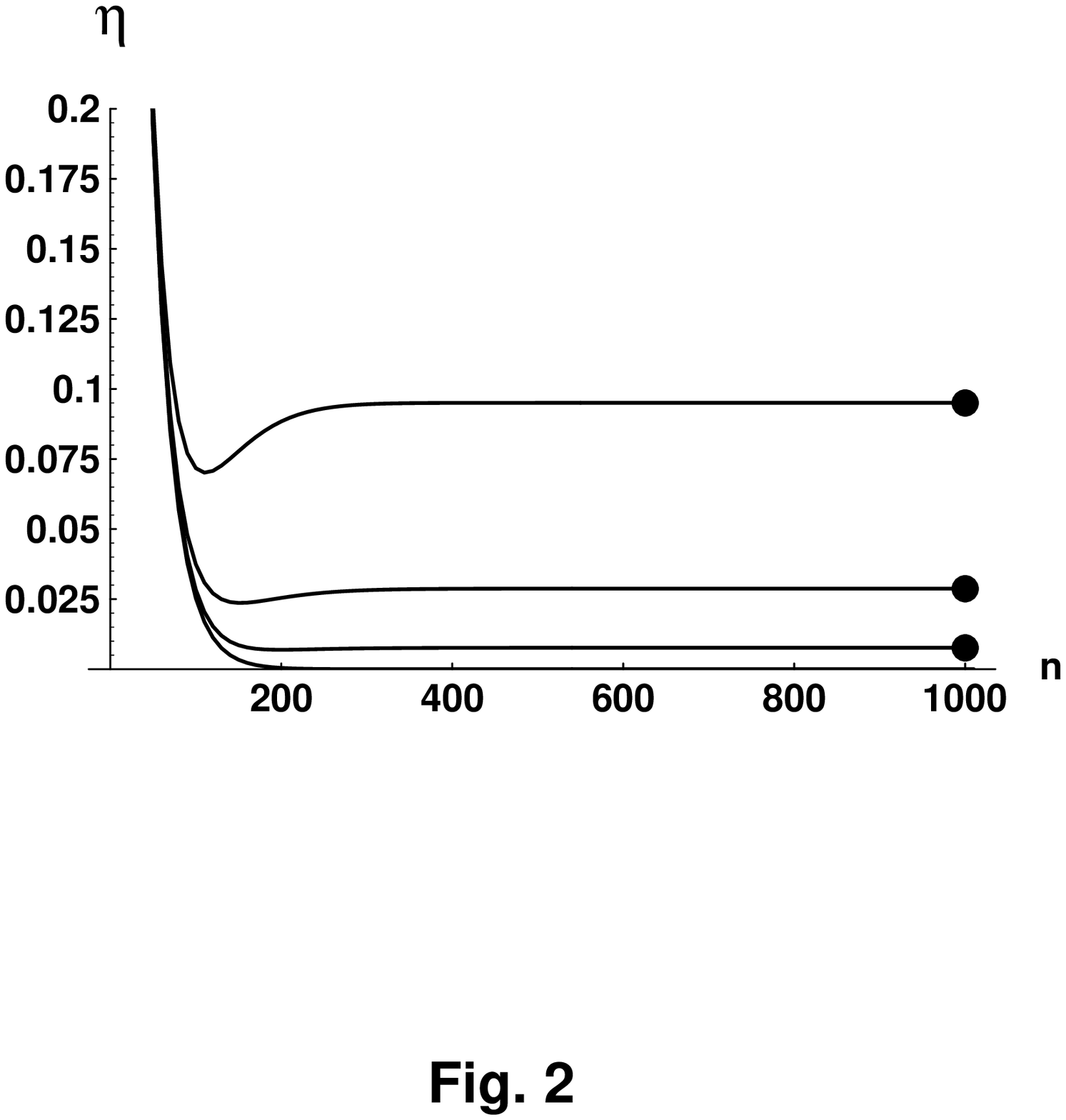}

\vfill\eject 

\epsffile{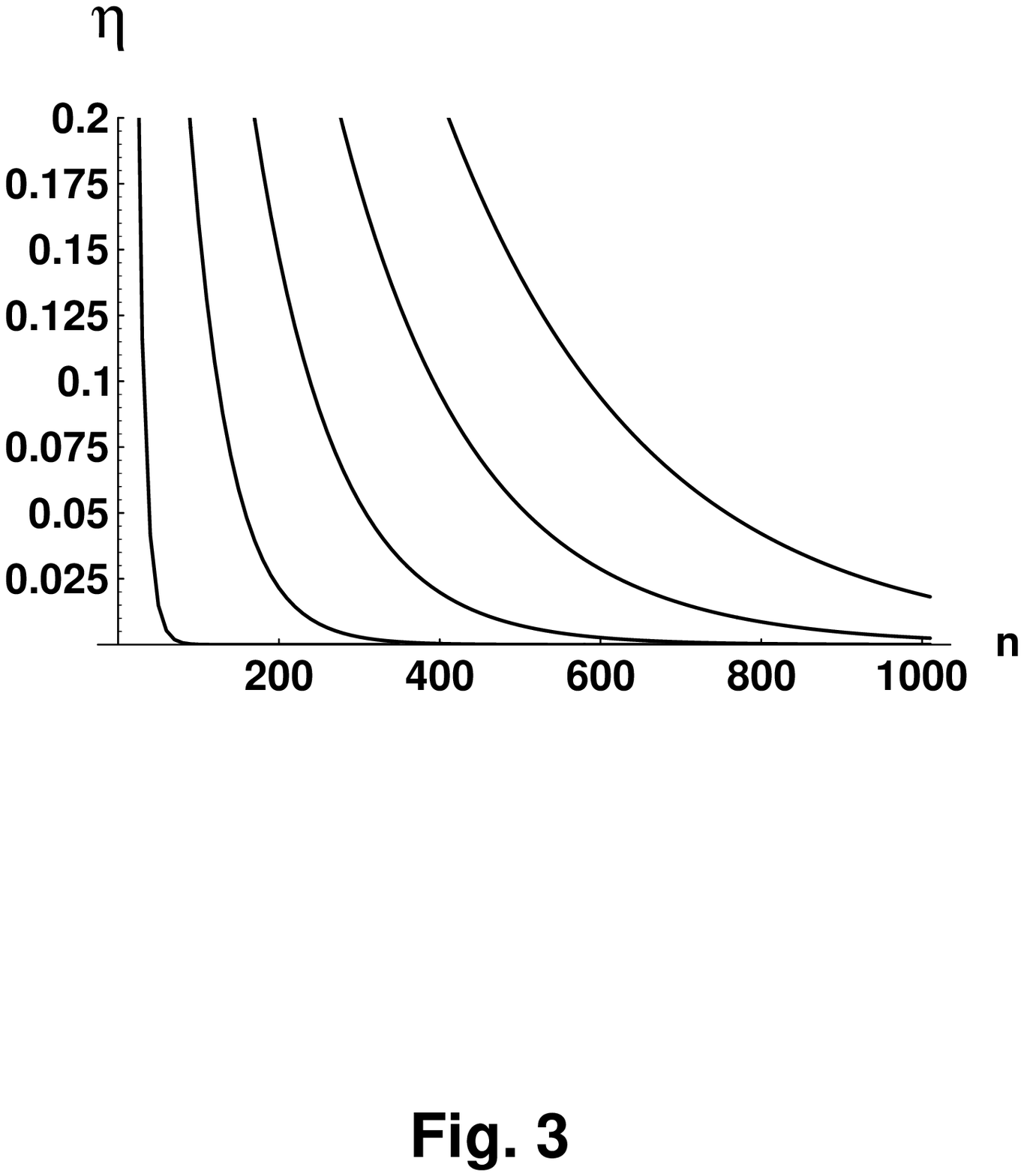}

\vfill\eject 

\epsffile{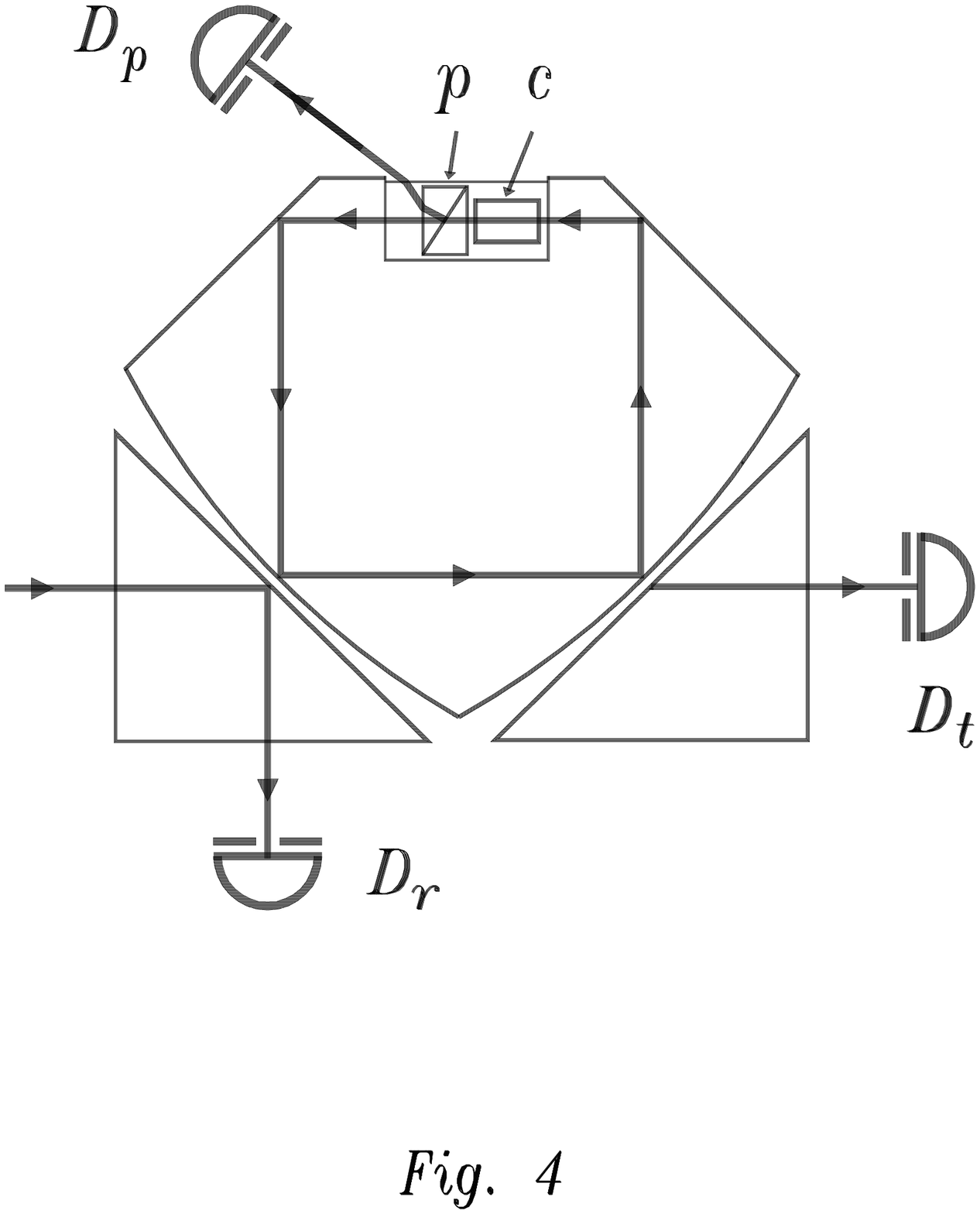}

\end{document}